\documentclass[aps,pra,amssymb, amsmath,nobibnotes,nobibnotes,reprint,superscriptaddress,
onecolumn, floatfix,showpacs,showkeys,notitlepage,longbibliography]{revtex4-2}
\usepackage{graphics}
\usepackage{graphicx, graphics, color,epsfig}
\usepackage{revsymb4-2}
\usepackage{bm}
\usepackage{braket}
\usepackage{mathptmx}
\usepackage[colorlinks, linkcolor=blue, anchorcolor=blue, citecolor=blue]{hyperref}
\usepackage{hyperref}
\usepackage{dsfont}
\usepackage{amsbsy}
\usepackage{amssymb}
\usepackage{xcolor}
\usepackage{amsmath}

\usepackage{orcidlink}

\begin{document}

\title{Optical forces on atoms subject to higher-order Poincar{\'e} vortex modes}

\author{Smail Bougouffa \orcidlink{0000-0003-1884-4861}\thanks{corresponding author}}
\email{sbougouffa@imamu.edu.sa and sbougouffa@hotmail.com}
\affiliation{Department of Physics, College of Science, Imam Mohammad Ibn Saud Islamic University (IMSIU), P.O. Box 90950, Riyadh 11623, Saudi Arabia}

\author{ Mohamed Babiker\orcidlink{0000-0003-0659-5247}}
\email{m.babiker@york.ac.uk }
\affiliation{School of Physics, Engineering and Technology, University of York,
York, YO10 5DD, United Kingdom}

\date{\today}

\begin{abstract}
The interaction of atoms with higher-order Poincar\'e optical vortex modes of order $m\geq 0$ is explored for light close to resonance with atomic dipole transitions.  It is well-known that atoms subject to optical vortex modes experience both translational and rotational forces acting on the atomic centre of mass, leading to atom dynamics and atom trapping.  Here we consider the optical forces on atoms immersed in general paraxial higher-order Poincar\'e optical vector modes.  The coupling to atoms gives rise to wide-ranging scenarios involving such modes in which any specific polarisation is within a spectrum of wave polarisation and all the interactions are treatable within a single formulation.  We show that this gives rise to a variety of physical situations, governed by the mode order $m$, the polarisation represented by the angular coordinates of the mode on the surface of the unit Poincar\'e sphere, the atomic transitions involved, and their selection rules. We present the analytical steps leading to the optical forces on sodium atoms and display their variations in various situations.

\end{abstract}

\keywords{Optical forces, Dipole interactions, Optical vortex beams, atom dynamics}

\pacs{ 37.10.De; 37.10.Gh }

\maketitle

\section{Introduction}\label{sec1}

It is by now well-established that any paraxial higher-order optical vortex vector mode of any order $m\geq 0$ can be represented by a unique point of spherical angular coordinates $(\Theta_P,\Phi_P)$ on the surface of the order $m$ unit Poincar\'e sphere and such modes are thus referred to as higher-order Poincar\'e modes. For general points $(\Theta_P,\Phi_P)$ the Poincar\'e modes are in general vector modes characterised as linear combinations of scalar modes.  They become pure scalar modes strictly at the north pole and the south pole where $\Theta_P=0,\pi$, respectively for all mode orders $m$.  Optical vortex modes, in general, have been highlighted as useful in several reports, particularly those involving laser physics and quantum optics, \cite{milione2011, milione2012higher,volpe2004,maurer2007, Zhan2009, liu2014,naidoo2016, Holmes_2019, GALVEZ202195,chen2020,liang2024,fickler2024} and in a variety of applications, notably as a means for the realisation of quantum information processing \cite{calvo2006quantum,kok2010introduction,gu2023metasurfaces}, in quantum optomechanics \cite{habraken2010geometric,chang2010cavity}, in atom trapping and cooling \cite{rosales2018review,santamato2004photon, otte2020optical}, in topological photonics \cite{xu2021polarization}, in optical tweezers \cite{bhebhe2019demand, padgett2011tweezers}, in biophotonics and imaging \cite{shi2017deep}, and communication and information transfer \cite{liu2022generation,luo2023full}.
Not surprisingly, some effort has also been directed towards finding out the significant roles scalar vortex modes, in general, have in their interaction with atoms  \cite{franke2008advances,forbes2016creation,leach2002measuring,chen2020,zhao2023pseudo,adrian2024,lembessis2024}.

This article is concerned with the optical forces on atoms when subject, not just to scalar vortex modes, but when subject to general higher-order vector Poincar\'e  modes. We focus on interactions associated with dipole transitions in Na. Steady-state forces act on the atomic center of mass that control the motion of atoms in such light fields and we aim to determine
their characteristics across the wide mode spectrum provided by the general higher-order Poincar\'e optical vortex modes. We show that, in general, the optical forces acting on the atoms are not just space- and time-dependent, but because of the Doppler effect, they also depend on the velocity ${\bf v}({\bf r},t)$ of the atomic center of mass as the atom moves in response to these forces within the light fields. We develop the formalism leading to the velocity and space-dependent forces, but in order to explore the basic characteristics of these forces we shall focus on their Doppler-free forms by setting ${\bf v}=0$.  As we shall see, this reveals interesting features of the forces which depend, besides the type of vortex mode and atomic parameters, on the optical polarisation and the order of the Poincar\'e mode in question. The full dependence of the forces on space-time with the inclusion of the Doppler effect is best incorporated within the equations of motion. Here we shall not proceed to the  solutions of the equations of motion as they are complex, numerically-intensive and require the specification of initial conditions leading to dynamics and atomic trajectories within the light fields.

This paper is organized as follows. In section \ref{sec2} we set out the analytical forms of the electromagnetic fields of higher-order Poincar\'e modes of order $m$. Each mode is assumed to have a radius at focus at least equal to, or larger than the wavelength of the light  ($w_0\geq\lambda$) so that the axial field components are negligibly small (and so are dropped) compared to the transverse components.  The transverse fields are then normalised in terms of the applied power $\cal P$ needed to generate the vector mode as a linear combination of two oppositely circularly polarised Laguerre-Gaussian modes traveling along the $z$ axis.    
With the case of sodium D2 transitions in mind we deal in section \ref{sec3} with the derivation of the optical forces exerted on a Na atom 
by a general order $m$ Poincar\'e light mode at near-resonance with the Na dipole transitions. These forces are shown to be critically dependent on the dipole selection rules, which lead to different contributions from the fields associated with the two-component fields forming the vector Poincar'e vortex mode. Section \ref{sec4} is concerned with the displays of the radial variations of the force components on the focal plane for different orders $m$ and for representative points $(\Theta_P,\Phi_P)$ on the surface of order $m$ unit Poincar\'e sphere. We also highlight the roles of the Gouy and curvature phase functions on the nature of the forces on either side of the focal plane which indicate sudden changes in the force components as the atom traverses the focal plane.  Both kinds of force, namely the scattering and the dipole forces are necessary for determining the motion of such sodium atoms immersed in the vector Poinacar\'e modes of order $m$.
Section \ref{sec5} contains our comments and final conclusions.

\section{Higher Poincar{\'e} modes}\label{sec2}
A typical paraxial higher order Poincar\'e mode is formed as an inseparable superposition of two oppositely circularly polarised optical vortex modes, one with the right-hand circularly polarised optical vortex of winding number $m$ and the second is left-handed circularly polarised vortex with winding number $-m$ and both have the same amplitude function and traveling along the common axis with the same axial wavevector $k_z$  as indicated by the spatial phase $e^{ik_zz}$ and frequency $\omega$.  
For this combination to be a Poincar\'e mode the two components, labeled 1 and 2  are weighted by Poincar\'e functions ${\cal U}_P$ and ${\cal V}_P$, respectively \cite{babiker2024}. These ensure that every point on the surface of the order $m$ unit Poincar\'e sphere corresponds to a Poincar\'e vortex mode with a well-defined polarisation at every point as either right circularly-polarised, right or left elliptically-polarised, radially- or azimuthally-polarised and left-circularly polarised.
 
The electric and magnetic  fields of this general Poincar\'e mode are as follows 
\begin{equation}
{\bf E}=({\bf E}_1+{\bf E}_2);\;\;\;\;\;\;{\bf B}=({\bf B}_1+{\bf B}_2). \label{one}
\end{equation}
Here ${\bf B}_i$ and ${\bf E}_i$, with $i=1,2$, are associated with the order $m$ Poincar\'e mode which 
have the following analytical forms \cite{babiker2024}. For $i=1$ the fields are  given in terms of the cylindrical coordinates $(r, \phi, z)$ as follows
\begin{eqnarray}\label{P10}
{\bf E}_1&=c\left\{ik_z({\bf {\hat e_x}}-i{\bf {\hat e_y}})\right\} {\cal U}_P e^{i m\phi}{\tilde {\cal F}}_{m, p}((\rho,z))e^{ik_zz-i\omega t}\nonumber\\
{\bf B}_1&=\left\{ik_z({\bf {\hat e_y}}+i{\bf {\hat e_x}})\right\} {\cal U}_P e^{i m\phi}{\tilde {\cal F}}_{m, p}((\rho,z))e^{ik_zz-i\omega t}
\label{upee}
\end{eqnarray}
and for the fields ${\bf E}_2$ and ${\bf B}_2$, the expressions are as follows
\begin{eqnarray}\label{P11}
{\bf E}_2&=c\left\{ik_z({\bf {\hat e_x}}+i{\bf {\hat e_y}})\right\} {\cal V}_P e^{-i m\phi}{\tilde {\cal F}}_{m, p}((\rho,z))e^{ik_zz-i\omega t}\nonumber\\
{\bf B}_2&=\left\{ik_z({\bf {\hat e_y}}-i{\bf {\hat e_x}})\right\} {\cal V}_P e^{-i m\phi}{\tilde {\cal F}}_{m, p}((\rho,z))
e^{ik_zz-i\omega t}.
\label{vpee}
\end{eqnarray}
Note that the fields do not include any longitudinal (axial) components as we are concerned with weakly focused paraxial modes for which the beam waist $w_0$ is equal to or larger than the wavelength $\lambda$ in which case the axial field components are negligible compared with the transverse ones. The well-known criterion for neglecting the longitudinal component was due to Lax et al \cite{lax1975}.  It amounts to the imposition of the condition $w_0k_z>>1$ where $w_0$ is the beam width at focus and $k_z=2\pi/\lambda$ the axial wave number. This implies that for strong focusing where the longitudinal field component becomes substantial, we must have the beam waist of the order $w_0\approx 0.16\lambda$.  In this paper, we consider beam widths much larger than $0.16\lambda$ so that the longitudinal components are indeed negligible. In the field expressions above, the function
${\tilde {\cal F}}_{m, p}$ is the paraxial mode profile function of winding integer number $m \geq 0$ and integer radial number $p\geq 0$. We have expressed the right and left circular polarisations in terms of the Cartesian unit vectors $\mathbf{ \hat{e}_{i}} (i=x,y)$.  The factors   ${\cal U}_P$ and ${\cal V}_P$, which appear in the field expressions, are Poincar\'e functions given by  \cite{babiker2024}
\begin{equation}\label{P3}
{\cal U}_P=\frac{1}{\sqrt{2}}\cos{\left(\frac{\Theta_P}{2}\right)}e^{-i\Phi_P/2}
;\quad
{\cal V}_P=\frac{1}{\sqrt{2}}\sin{
\left(\frac{\Theta_P}{2}\right)}e^{i\Phi_P/2},
\end{equation}
where $\Theta_P$ and $\Phi_P$ are the Poincar\'e angles, which are defined at every point on the surface of the unit Poincar\'e sphere \cite{babiker2024}, and the Laguerre-Gaussian (LG) amplitude function
$\tilde {\cal F}_{m, p}$  for  winding number $m$ and radial number $p$ which is
\begin{widetext}
\begin{equation}
{\tilde {\cal F}}_{m,p}(\rho,z)={\cal A}_0\frac{C_{|m|,p}}{\sqrt{1+z^{2}/z_{R}^{2}}}\left( \frac{\rho \sqrt{2}}{w_{0} \sqrt{1+z^{2}/z_{R}^{2}}}\right)^{|m|}
\exp\left[\frac{-\rho^{2}}{w_{0}^{2}(1+z^{2}/z_{R}^{2})}\right]L_{p}^{|m|}\left\{\frac{2\rho^{2}}{w_{0}^{2}(1+z^{2}/z_{R}^{2})}\right\}e^{i\xi(\rho, z)},
\label{eq:refname3}
\end{equation}
\end{widetext}
where ${\cal A}_0$ is a normalisation factor, to be determined in terms of the power ${\cal P}$. The phase function $\xi(\rho, z)$  includes the Gouy and the curvature phases 
\begin{equation}
\xi(\rho, z)=- (2p+|m|+1)\arctan\left(\frac{z}{z_{R}}\right)+\frac{k_zz\rho^{2}}{2(z^2+z_{R}^{2})}.
\label{xi}
\end{equation}
Here $w_0$ is the radius of the mode at focus, $z_{R}=w_0^2k_z/2$ is the Rayleigh range, $C_{|m|,p}=\sqrt{p!/(p+|m|)!}$ and $L_{p}^{|m|}$ the associated Laguerre polynomial. Note that ${\tilde{\cal F}}$ only depends on $|m|$, hence the Poincar\'e modes of order $m$, are understood as either LG$_{m,p}$ or LG$_{-m,p}$ modes. 

The normalisation factor ${\cal A}_0$ can be expressed in terms of the applied power $\cal P$ by integrating the z-component of the Poynting vector of the whole mode over the beam cross-section \cite{babiker2024super}
\begin{equation}\label{P4}
{\cal P}=\frac{1}{2\mu_0}\int_0^{2\pi}d\phi\int_0^{\infty}|({\bf E}^*\times{\bf B})_z|\rho d\rho.
\end{equation}
 On substituting for the fields and integrating we obtain the normalisation factor ${\cal A}_0$ 
\begin{equation}
{\cal A}_0^2=  \frac{4\mu_0 {\cal P}}{\pi c k_z^2w_0^2}  
\end{equation}

The normalised electric and magnetic fields of the paraxial Poincar\'e modes of the Laguerre-Gaussian form have thus been fully specified and we are in a position to consider their coupling to atoms.

\section{Optical forces on atoms}\label{sec3}
\subsection{Scattering and dipole forces}

The essential elements of the interaction of an atom with a  Poincar\'e mode at near-resonance are described in terms of the two-level neutral atom in which the ground and excited states are denoted $\{\ket{g}, \ket{e}\}$ whose energy levels are $\mathcal{E}_1$ and $\mathcal{E}_2$, respectively. The transition frequency is then $\omega_a=(\mathcal{E}_2- \mathcal{E}_1)/\hbar$. The interaction Hamiltonian involves the optical fields evaluated at the center of mass coordinate $\mathbf{r}$ and we shall focus on sodium atom electric dipole transitions for each of which the interaction Hamiltonian and the Rabi frequency ${\Omega}$ are defined as  
\begin{equation}\label{Q1}
 \hat{H}_{int}=-{\bf d}\cdot\mathbf{{E}};\;\;\;\;\;\;\;\;\;\Omega=\frac{|{\bf d}\cdot{\bf E}|}{\hbar},
\end{equation}  
where ${\bf d}=e\mathbf{q}$ is the electric dipole moment vector with $\mathbf{q}$ the internal position vector. Other parameters relevant to the optical forces on atoms are the detuning $\Delta$ and the spontaneous emission rate $\Gamma$ which are defined below.

The steady-state optical forces at near resonance with a two-level atom are Doppler forces which are present due to any form of the light field. These forces are well-known in the limit of moderate field intensity \cite{Domokos2003, mohammed2023, bougouffa2021quadrupole}. The total average force ${\bf F}$ due to the dipole interaction with an atom moving with velocity $\mathbf{v}=\dot{\mathbf{r}}$ is the sum of two parts
\begin{equation}\label{O1}
 {\bf F} (\mathbf{r},\mathbf{v}) ={\bf  F}_{sca} (\mathbf{r},\mathbf{v}) +{\bf F}_{dip} (\mathbf{r},\mathbf{v}),
\end{equation}
The first term is the scattering force due to the absorption and spontaneous emission of light by the moving atom via dipole transitions. The second is the dipole force, which acts to confine the atom to maximal or minimal intensity regions of the field. Both forces can be written in terms of the saturation function ${S}$
\begin{equation}
{S}=\frac{\Omega^2/2}{\Delta^2+\Gamma^2/4}.
\end{equation}

Here,  $\Omega(\mathbf{r})$ is the dipole Rabi frequency, $\Gamma$  is the atomic dipole transition rate and $\Delta(\mathbf{r},\mathbf{v})$ is the dynamic detuning which is a function of both the atomic position vector ${\bf r}$ and the velocity vector ${\bf v}={\dot{\bf r}}$. We have  $\Delta (\mathbf{r},\mathbf{v})=\Delta_{0} -\mathbf{v}\cdot \nabla \Psi(\mathbf{r}),$ where $\Delta_{0} =\omega -\omega_{a} $ is the static detuning, with $\omega$ the frequency of the applied light field and $\Psi({\bf r})$ is the phase. The second term in the dynamic detuning $\Delta$ is written $\delta =-\mathbf{v}\cdot \nabla \Psi (\mathbf{r})$ and arises because of the Doppler effect due to the atomic motion. The dipole force confines the atom to maximal or minimal intensity regions of the field, depending on the detuning $\Delta$ sign. 
We have the two forces
\begin{equation}
{\bf F}_{\text{sca}}=\frac{\hbar \Gamma}{2} \frac{S}{1+S} {\bf \nabla} \Psi;\;\;\;\;{\bf F}_{dip}=-\frac{\hbar\Delta}{\Omega}\frac{S}{1+S}{\bf \nabla}\Omega\label{forces}
\end{equation}
\begin{figure}
\includegraphics[width=0.35\linewidth,height=0.32\linewidth]{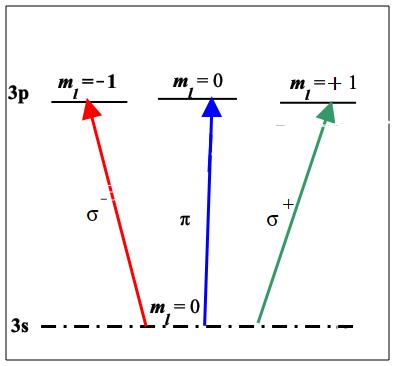}
\caption{Schematic figure showing the transitions representing the D2-line in sodium. The relevant transitions here are those satisfying the selection rule $\Delta m_{\ell}=\pm 1$.}
\end{figure}

Corresponding to the optical force is an optical potential which has the form
\begin{equation}\label{O4}
U_{dip} (\mathbf{r}) =\frac{\hbar \Delta}{ 2} \ln(1+S).
\end{equation}

For red-detuned light $\Delta _{0} <0$, the potential exhibits a (trapping) minimum in the high-intensity region of the beam which is detuned below resonance (where $\omega<\omega _{a}$). For blue detuning $\Delta _{0} >0$, the trapping process takes place in the low-intensity (dark) regions of the field.  Furthermore, in many experimental situations and when the detuning is large and is such that $(\Delta\gg \Omega)$ and $(\Delta\gg \Gamma)$  then the optical potential can be approximated by
\begin{equation}\label{O5}
    U_{dip} (\mathbf{r}) \approx \frac{\hbar}{\Delta} \Omega^2.
\end{equation}

The general forms of the optical forces stated above, which are applicable in an arbitrary light field, form the basis of exploring their characteristics specifically for Na atoms immersed in the paraxial higher-order Poincar\'e modes. As pointed out above and stated at the outset, the optical forces on the atom are velocity- and space-time dependent.  We have seen that the velocity dependence enters through the dynamic detuning $\Delta=\Delta_0-\mathbf{v}\cdot \nabla \Psi$, but in order to explore the basic characteristics of these forces we shall focus on their Doppler-free forms by setting $ \Delta=\Delta_0$.

\subsection{Selection rules}
For transitions between two specific levels the atom is described as a two-level system, so that the dipole moment operator can be written in terms of ladder operators as $\hat{\bm{d}}=\bm{d}_{\alpha\beta}\hat{b}+\bm{d}_{\alpha\beta}^{*}\hat{b}^{\dag}$ where $\alpha$ and $\beta$ take the values $1,2$. Here, the induced atomic dipole matrix elements between the two levels are denoted by $\bm{d}_{\alpha\beta}=\bra{\alpha}\hat{\bm{d}}_{\alpha\beta}\ket{\beta}$, and the atomic level lowering and raising operators are represented by $\hat{b}$ and $\hat{b}^{\dag}$, respectively. The atom is assumed to have zero permanent atomic dipole moments  $(\bm{d}_{11}=\bm{d}_{22}=0)$. 

The total electric field ${\bf E}={\bf E}_1+{\bf E}_2$ of the general  Poincar\'e mode deduced from inspection of Eqs. (\ref{one},\ref{P10}, \ref{P11}) can now be written as follows in Cartesian coordinates
\begin{equation}\label{Q2}
    \mathbf{ E}(\mathbf{r},t)=\sum_{i}\mathbf{ \hat{e}_{i}}E_i({\bf r},t),
\end{equation}
where $\mathbf{ \hat{e}_{i}} (i=x,y,z)$ are the Cartesian unit vectors and $E_i$ are the Cartesian components of the electric field and can be written as
\begin{equation}\label{Q3}
E_i({\bf r})= u_i({\bf r})e^{i sk_zz}e^{-i\omega t},
\end{equation}
where  $u_i(\mathbf{r})$ are the amplitude functions of the Cartesian $i^{th}$ total electric field component $E_i=E_{1i}+E_{2i}$ of the general Poincar\'e mode, which emerge straightforwardly as follows 
\begin{eqnarray}
u_x(\mathbf{r}) & = & ik_zc\left( {\cal U}_P e^{i m\phi}+{\cal V}_P e^{-i m\phi}\right) {\tilde {\cal F}}_{m, p}(\rho,z)\label{e13}\\
u_y(\mathbf{r}) & = & k_zc\left( {\cal U}_P e^{i m\phi}-{\cal V}_P e^{-im\phi}\right){\tilde {\cal F}}_{m, p}(\rho,z). \label{e14} 
\end{eqnarray}

 The general Poincar\'e mode of order $m$ is specified by its Poincar\'e angles $\Theta_P,\Phi_P$, frequency $\omega$, the Laguerre-Gaussian function with its azimuthal and radial numbers $m$ and $p$, and the direction of propagation index $s$. 

The interaction of the atom is characterized by the set of Rabi frequencies each of which depends on the dipole moment vector and the electric field vector of the mode, evaluated at the atomic centre of mass coordinate ${\bf r}$. 
For application to Na atoms,  the dipole active transitions of interest occur from the state $\ket{L=0,m_{\ell}=0}$ to the state $\ket{L=1,m_{\ell'}}$ 
\cite{Bougouffa2020a, Bougouffa2020, babiker2018atoms, lembessis2013enhanced, bransden2003physics, bougouffa2023absorption}. In the application of the selection rules, we distinguish different cases based on the values of $m_{\ell'}$ that satisfy the selection rules.
  For $m_l' = 0$, we can calculate the dipole moment tensor \cite{Bougouffa2020a, Bougouffa2020, Barnett2022, ficek2014quantum,le2018chiral, metcalf1999laser}. As a result, the dipole matrix elements have components of $(d_x=d_y=0, d_z=d_{eg})$. This simplifies the product ${\bf d.E}$, which is, apart from a minus sign, is the interaction energy.  This has the following form:
\begin{equation}\label{e20a}
\left({\bf d}\cdot{\bf E}\right)_0={d_{eg}}{u_z} e^{i(k_z z)}.
\end{equation}
However, since the axial field amplitude $u_z$ is assumed to be very small for beams with large $w_0\geq\lambda$, this interaction can be regarded as negligibly small.
  For $m_l=\pm 1$, the dipole matrix element components are $(d_x=d_{eg}, d_y=\pm i d_{eg}, d_z=0)$. Therefore, the product  ${\bf d.E}$ is now such that:
 \begin{equation}\label{e21}
 \left({\bf d}\cdot{\bf E}\right)_{\pm} ={d_{eg}}\Big [u_{x}\pm iu_{y} \Big]e^{i(k_z z)}.
 \end{equation}    
 Substituting for $u_{x}$ and $u_{y}$ from Eqs.(\ref{e13}) and (\ref{e14}), we have 
\begin{equation}
     \left({\bf d}\cdot{\bf E}\right)_{+}={2k_zcd_{eg}}{\cal U}_P{\tilde {\cal F}}_{m, p}e^{im\phi};\;\;\;\;\;
     \left({\bf d}\cdot{\bf E}\right)_{-}={2k_zcd_{eg}}{\cal V}_P{\tilde {\cal F}}_{m, p}e^{-im\phi}.
     \label{plusminus}
\end{equation}
We see that the interaction energies corresponding to the two different transitions differ only in that they involve different Poincar\'e functions ${\cal U}_P$ and ${\cal V}_P$.  The former is associated with the right-hand circularly-polarised term and the latter with the left-circularly-polarised term which first appeared in the electric field expressions of the Poincar\'e mode, Eqs.(\ref{upee}).and (\ref{vpee}).

\subsection{Rabi frequencies and total optical forces}
It is clear that this physical system involves the participation of two separate transitions labeled by the subscripts $+$ and $-$, with corresponding dipole interaction Hamiltonians given by Eqs(\ref{plusminus}).  There are thus two Rabi frequencies denoted $\Omega_{\pm}$ which are given by 
\begin{equation}
\Omega_+^2({\bf r})= \left| \frac{{\bf d}\cdot{\bf E}}{\hbar}\right|_+^2=\Omega_{m,p}^2({\bf r})\cos^2{\left(\frac{\Theta_P}{2}\right)};\;\;\;\;\Omega_-^2({\bf r})= \left| \frac{{\bf d}\cdot{\bf E}}{\hbar}\right|_-^2=\Omega_{m,p}^2({\bf r})\sin^2{\left(\frac{\Theta_P}{2}\right).}
\end{equation}
where $\Omega_{m,p}^2$ is given by 
\begin{equation}
  \Omega_{m,p}^2({\bf r})=  \left(\frac{2k_z^2c^2{d}_{eg}^2}{\hbar^2}\right)|{\tilde {\cal F}}_{m, p}(\bf r)|^2.
\end{equation}
Note that the Rabi frequencies depend only on the Poincar\'e angle $\Theta_p$ and have the same values for arbitrary $\Phi_P$. 

We are now in a position to derive the forms of the scattering forces and the dipole forces associated with each transition and proceed to add these forces to obtain the total force of each kind as the sum of contributions due to the two transitions.  In doing so we shall assume that the upper states have the same spontaneous emission rate $\Gamma$ and the same transition probability.

The scattering forces require the evaluation of phase function $\Psi_{\pm}$  associated with the two transitions and these can be deduced from Eqs. (\ref{plusminus}).  We have 
\begin{equation}
   \Psi_{\pm} =k_zz\pm m\phi+\xi(\rho,z).
\end{equation}
Since we have two different Rabi frequencies associated with the two transitions we now have two scattering forces which are functions of the atomic position vector ${\bf r}=(\rho,\phi,z)$  
\begin{equation}
{\bf F}_{sca\pm}({\bf r})=\frac{\hbar \Gamma}{2} \frac{S_{\pm}}{1+S_{\pm}} {\bf \nabla} \Psi_{\pm}({\bf r}),
    \label{scat+-}
\end{equation}
where $S_{\pm}$ are given by
\begin{equation}
{S}_{+}({\bf r})=S_{m,p}\cos^2{\left(\frac{\Theta_P}{2}\right)}, \quad {S}_{-}({\bf r})=S_{m,p}\sin^2{\left(\frac{\Theta_P}{2}\right)};\;\;\;S_{m,p}({\bf r})=\frac{\Omega_{m,p}^2/2}{\Delta^2+\Gamma^2/4 .}\label{ess+-}
\end{equation}

Evaluating the gradient of the phases $\Psi_{\pm}$ we have
\begin{equation}
{\bf \nabla} \Psi_{\pm} =\frac{\partial\xi(\rho,z)}{\partial\rho}{\bf \hat\rho}\pm\frac{m}{\rho}{\bf \hat\phi}+\left(k_z+\frac{\partial\xi(\rho,z)}{\partial z}\right){\bf {\hat z}}\label{Psi+-},
\end{equation}
where carets denote unit vectors. Evaluating the gradients ${\bf \nabla} \Psi_{\pm}$ using Eq.(\ref{Psi+-}) then leads us to the two individual forces ${\bf F}_{sca\pm}$.  We have assumed that both transitions are equally probable so, as we describe below, the total scattering force on the atom is the vector sum of the two scattering forces.

Turning now to the two dipole forces, we have  
\begin{equation}
{\bf F}_{dip \pm}=-\frac{\hbar\Delta}{\Omega_{\pm}}\frac{S_\pm}{1+S_\pm}{\bf \nabla}\Omega_{\pm}=-\frac{\hbar\Delta}{\Omega_{\pm}}\frac{S_\pm}{1+S_\pm}\left(\frac{\partial \Omega_{\pm}}{\partial \rho}\hat{\rho}+\frac{\partial \Omega_{\pm}}{\partial z}\hat{z} \right).
 \label{dip+-}
\end{equation}
Only the analytical evaluations of the components of ${\bf \nabla}\Omega_{\pm}$ are needed to determine the dipole forces.

As indicated above, the total scattering and dipole forces are the vector sums of the individual components
\begin{equation}
{\bf F}_{sca}={\bf F}_{sca +}+{\bf F}_{sca -};\;\;\;\;\;{\bf F}_{dip}={\bf F}_{dip +}+{\bf F}_{dip -}.
\end{equation}
Although the combined scattering force ${\bf F}_{sca}$ and the combined dipole force ${\bf F}_{dip}$ act on an atom at the same space-time leading to the atom dynamics, it is instructive to study the characteristics of the two types of force separately.  


\section{Application to sodium D2 transitions}\label{sec4}
In order to explore the characteristics and variations with parameters of the total forces acting on an atom due to interaction with higher-order Poincar\'e modes we focus on the case of the sodium atom and its transitions between the states 3S$_{1/2}$ and 3P$_{3/2}$.  The energy level diagram shown in Fig. 1 displays the specific transitions in question which are those satisfying $\Delta m_l=\pm 1$. 
The transition wavelength has the value $\lambda=589$ nm and the spontaneous emission rate is $\Gamma=6.15\times 10^7 Hz$, corresponding to transitions within the D2 line of the sodium atom. 

\begin{figure}
 \includegraphics[width=0.35\linewidth,height=0.3\linewidth]{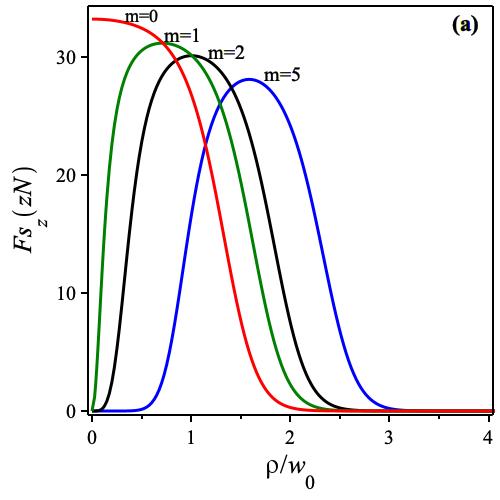}~\includegraphics[width=0.35\linewidth,height=0.31\linewidth]{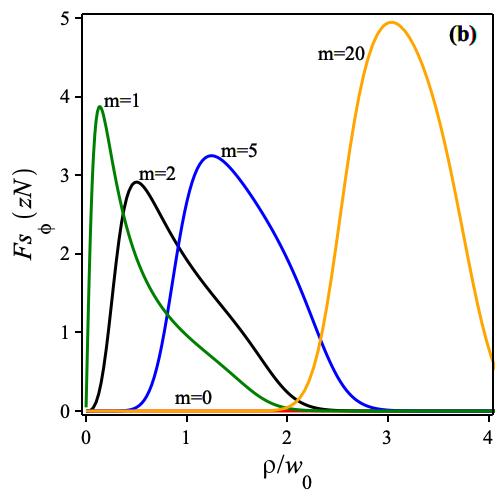}
\caption{Variations with the radial coordinate $\rho$ (in units of the beam waist $w_0$) of the components of the scattering force due to Laguerre-Gaussian doughnut (i.e. all have $p=0$) Poincar\'e modes on the focal plane $z=0$.  The parameters are such that the winding number is $m$, the radial number is $p=0$ throughout; the detuning is $\Delta_0=10\Gamma$; the power is ${\cal P} = 2.5\mu W $; the two transitions contributing to the forces in Na have the same wavelength $\lambda=589$ nm and the beam radius at focus is $w_0=5\lambda$.  
Here we have 
(a) the axial $(z)$ component of the scattering force, and (b) the azimuthal component (i.e. the ${\bf {\hat \phi}}$ component) of the scattering force.  The colour code in the plots is such that $m=0$ are represented by the red solid curve in (a); the $m=1$ by the green solid curves; the $m=2$ by the black solid curves; the $m=5$ by the blue solid curve, and the $m=20$ by the yellow solid curve. Note that the radial component of the scattering force is zero on the focal plane. The vertical axis in this figure (and also in subsequent figures) is in units of zepto Newton $zN=10^{-21} N$. }\label{Fig2}
\end{figure}

We focus on Poincar\'e modes of the Laguerre-Gaussian (LG) type and consider specifically doughnut modes for which the winding number is $m$, but the radial number is $p=0$.   In Figs. (\ref{Fig2}), we present the in-plane variations of the components of the scattering force at the focal plane $z=0$ as functions of $\rho$. The plots concern Poincar\'e modes of different order $m$ but all have $\Theta_P=0$,  $\Phi_P=0$; i.e. they all lie at the north pole point of every order $m$ unit sphere. The considerable changes that arise for general modes represented by other points $(\Theta_P,\Phi_P)$ on the surface of the order $m$ Poincare sphere are discussed in Fig.\ref{Fig4} below. Figs. (\ref{Fig2}) show that the axial component of the scattering force is much larger than the azimuthal component, while the radial component remains zero. Additionally, the peak of the axial component decreases as $m$ increases and shifts further from the origin. But the peak of the azimuthal component is larger for $m=1$ and decreases for $m=2$ then increases with large values of $m>2$.

\begin{figure}
 \includegraphics[width=0.4\linewidth,height=0.32\linewidth]{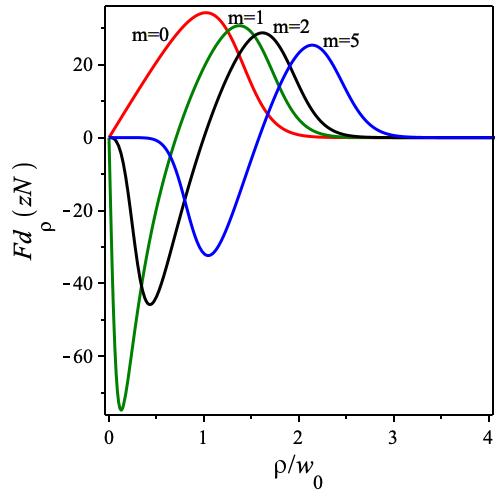}
 \caption{Variations with the radial coordinate $\rho$ (in units of the beam waist $w_0$) of the radial (only) component of the dipole force due to Laguerre-Gaussian doughnut Poincar\'e modes on the focal plane $z=0$.  The parameters and the colour codes of the different curves are the same as in Fig.\ref{Fig2}. Again, all modes of order $m$ lie at the north pole point of the respective unit sphere of order $m$. There are notable changes that arise for modes represented by points $(\Theta_P\neq 0,\Phi_P)$ as explained in Fig.4 below.}\label{Fig3}
\end{figure}
Figure (\ref{Fig3}) displays the variations with the scaled radial coordinate $\rho/w_0$ of the radial component of the dipole force at focal plane $z=0$ for different Poincar\'e orders $m=0,1,2,5$  and with other parameters as in previous figures. The variations are dominated by the gradient of the Rabi frequency which, as for the gradient of the field intensity, has the characteristic form in which there is a region where the gradient is negative, followed by a minimum. It then goes through zero followed by positive variations to a maximum and finally an exponential decay due to the variations of the amplitude of the Laguerre-Gaussian function.  The azimuthal and axial components of the dipole force are both zero at $z=0$. 
\begin{figure}[h]
 \includegraphics[width=0.31\linewidth,height=0.3\linewidth]{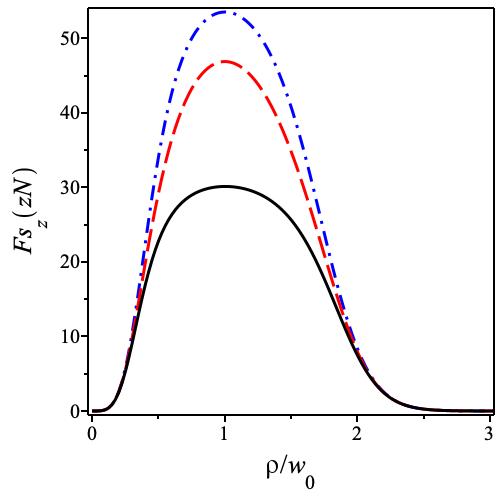}
 ~\includegraphics[width=0.31\linewidth,height=0.3\linewidth]{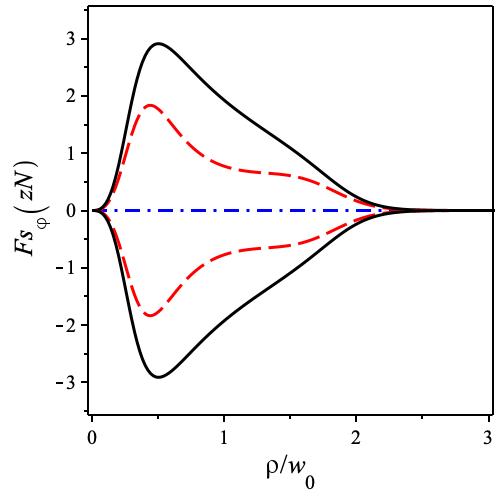}
 ~\includegraphics[width=0.31\linewidth,height=0.3\linewidth]{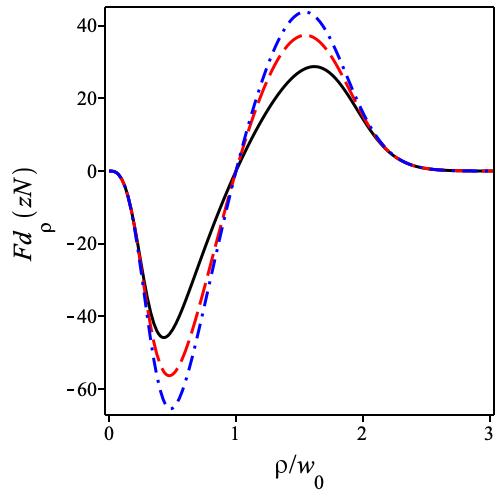}
\caption{Variations with the scaled radial coordinate $(\rho/w_0)$ of the axial component of the scattering force (left panel), the azimuthal component of the scattering force (middle panel), and the radial component of the dipole force (right panel).  The plots are for the second order for which  $m=2$ and $p=0$, evaluated at the focal plane and also at representative points on both the northern and southern hemispheres of the Poincar\'e sphere: $\Theta_P=0, \frac{\pi}{4}, \frac{\pi}{2}, 3\pi/4$ and $\pi$, with $\Phi_P=0$ throughout. The power is ${\cal P} = 2.5\mu W $; the detuning is $\Delta_0=10\Gamma$; the wavelength of the light is $\lambda=589$ nm and the radius of all modes at focus is $w_0=5\lambda$.  Here the colour code is such that the $\Theta_P=0,\pi$ is represented by the black solid curve, the $\Theta_P=\frac{\pi}{4},3\pi/4$ by the red dashed curve, and the $\Theta_P=\frac{\pi}{2}$ by the blue dash-dotted curve. Note that in the left panel and the right panel, the curve for $\Theta_P=3\pi/4$ coincides with that for $\pi/4$ and the $\Theta_p=\pi$ coincides with $\Theta_P=0$. Note also that other force components are identically zero on the focal plane.}
\label{Fig4}
\end{figure}

Figure (\ref{Fig4}) shows the changes that occur for representative Poincar\'e modes in the case of $m=2$ which lie along the longitude, defined by $\Phi_P=0$ linking the north pole $\Theta_P=0$ of the $m=2$ Poincar\'e sphere to its south pole $\Theta_P=\pi$,  specifically for the Poincar\'e  modes at $\Theta_P=0, \pi/4, \pi/2,3\pi/4,\pi$. The relevant parameters are the same as those in previous figures. The plots here indicate that as $\Theta_P$ increases, the maximum magnitudes of the axial dissipative force component and the radial component of the dipole force increase. The radial component of the dipole force follows the variations of the gradient of the intensity which is negative initially, passes through a minimum, then passes through zero at maximum intensity, and finally decreases due to the exponential decay with radial distance in the mode amplitudes.  On the other hand, the peaks of the azimuthal component decrease with increasing $\Theta_P$ and have the feature that the variations in the north and south hemispheres are equal but opposite in sign. 
\begin{figure}
\includegraphics[width=0.45\linewidth,height=0.33\linewidth]{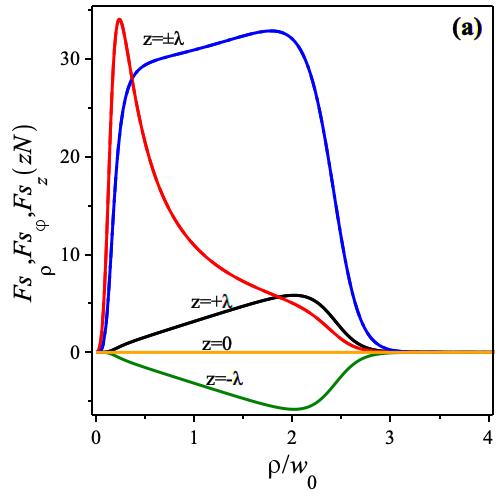}
\includegraphics[width=0.45\linewidth,height=0.33\linewidth]{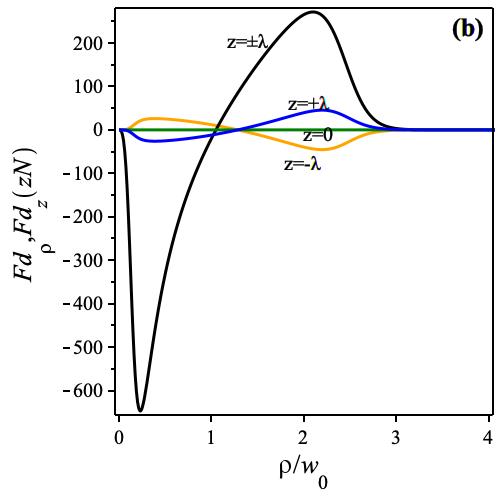}
\caption{Variations of the components of the scattering and dipole forces of the $m=2$ for two values of $z=\pm\lambda$ on the axis on either side of the focal plane.  The idea is to explore the effects of the Gouy and the curvature phases across the focal plane.  The other parameters are the same as those in previous figures, but $\Theta_P=0$ and $\Phi_P=0$.
(a) The red curve represents the variation with $\rho$ of the azimuthal component and the blue curve represents the axial component, both shown at $z=\pm \lambda $. The black, yellow, and green curves represent the radial component at $z=+\lambda$,  $z=-\lambda$, and $z=0$, respectively. (b) The black curve represents the radial component at $z=\pm\lambda$. The blue, green, and yellow curves represent the axial component at $z=+\lambda$, $z=0$, and $z=-\lambda$, respectively.}\label{Fig7}
\end{figure}

In general, the effects of the Gouy and curvature phase come into play for small beam widths $w_0$ of the order of the wavelength. 
 Thus, It is interesting to examine how they affect the optical forces exerted on atoms due to the Poincar\'e modes. The strongest variations occur on crossing the focal plane. Figure \ref{Fig7} deals with the case $m=2$ and displays the force components at axial positions $z=+\lambda$ to the positive side of the focal plane on the axis and at $z=-\lambda$ on the negative side.  The figure shows that the only component that changes significantly is the azimuthal component which is zero on the focal plane at z=0, but displays the same magnitude variations differing only in sign between the two axial positions.  
 
\section{Conclusions}\label{sec5}
We have set out to determine as a first step the analytical forms of the optical forces acting on sodium atoms coupled to a single general higher-order Poincar\'e mode of the Laguerre-Gaussian type propagating along the positive +z-axis and for which the winding number is $m$. Still, the radial number $p$ is zero.  We have focused in particular on the D2 transitions and explained how the transition selection rules determine the combined form of the interaction Hamiltonian.  As a result, we have two Rabi-frequencies, which enter the two types of forces acting on the centre of mass of the atom.  The development of the analytical formalism turns out to be particularly convenient, but also transparent, in terms of the saturation functions, which are modulated by the Poincare angles.  They thus vary continuously with the mode position on the surface of any order $m>0$ Poincar\'e unit sphere.  

In applications to the sodium D2 transitions, we concentrated on the illustration of the components of the dissipative and dipole forces when the atom is at rest.  The examples we considered have thus set the scene for any applications involving the motion of the atom in such Poincar\'e modes in which case we have to employ the dynamic detuning $\Delta({\bf r,v})$ in the formalism, rather than $\Delta_0$, the static detuning.  Applications to atomic motion are beyond the scope of this paper as they require numerical solutions to complex equations of motion and require the imposition of initial conditions.  Our aim so far has been to explore the nature and characteristics of the forces and display their spatial variations and their changes with the location of the mode on the surface of the order $m$ unit Poincar\'e sphere.  

Our results indicate that one of the dominant force components is the axial dissipative force, as shown in Fig.\ref{Fig2} which, in traditional atom dynamics is a feature of scalar modes and normally requires the presence of an additional counter-propagating mode to counter-act this force component in order to confine the motion within specified regions of the light fields.  This device will still be needed when dealing with atom dynamics for atoms subject to the Poincare\'e vector modes where the situation also calls for counter-propagating Poincar\'e modes. The peak force of the axial component of the dissipative force is seen to change with the increasing order $m$  of the mode. The radial component of the dipole force is also a dominant force, as seen in Fig.\ref{Fig3}, and it also changes with the order $m$ of the mode. 

The azimuthal component of the scattering force for $m>1$ shows a general increase as $m$ increases.  However, the case $m=1$ exhibits a peak higher than that of the second order $m=2$ and we attribute this to the spin-to-orbit conversion mechanism such that the effective order is $m+\sigma$, even though the maxima occur at $\rho=w_0\sqrt{m/2}$, i.e. at the point of highest intensity where also the exponential factor $e^{-\rho^2/w_0^2}$ in the amplitude function ${\tilde{\cal F}}_{m,p}$ is largest.  The peak for the order $m=5$ is also lower than the $m=1$ peak.  The trend beyond this is an increase with increasing $m$, thereafter.

The variations with the mode Poincar\'e angular position $(\Theta_P, \Phi_P)$, which modulates the wave polarisation of the mode on the surface of the unit Poincar\'e sphere indicate clearly that considerable changes arise in the magnitudes of the optical forces on the atom, as shown in Fig.\ref{Fig4}. In particular, the axial component of the dissipative force increases from the initial position at the north pole along the longitude $\Phi_P=0$ continuously to the south pole.  The maximum height for the case considered where $m=2$ is when $\Theta_P=\pi/2$.  As $\Theta_P$ increases progressively within the south hemisphere towards the south pole, where $\Theta_P$ has the value of $\pi$, we get the same variations as for the north hemisphere.  This symmetry between the north and south hemispheres is not shown by the azimuthal ($\phi$) component which decreases as $\Theta_P$ increases reaching zero at $\pi/2$ and becomes negative thereafter in the south hemisphere, but with the same magnitude including the south pole.  The radial component of the dipole force follows the variations of the gradient of the intensity which is negative initially, passes through a minimum, then passes through zero at maximum intensity, and finally decreases due to the exponential decay with radial position in the mode amplitudes.

Finally, we examined the roles played by the Gouy and the curvature phases in the context of the optical forces across the focal plane. We chose a beam waist $w_0=\lambda$, which somewhat indicates tight focusing, but still satisfies the paraxiality criterion, in order to maximise the effects of the Gouy and curvature across the focal plane.  As shown in Fig.\ref{Fig7} we set out to evaluate the variations of the components of the two types of force at three axial positions $(z=-\lambda,0,+\lambda)$.  We find that the azimuthal component of the dissipative force and radial components of the dipole force display sudden changes across the focal plane.  This would influence the motion of the atoms in this region.

Having examined the nature and variations of the optical forces in this paper, the situation is now ripe for evaluating the motion of atoms in such Poincar\'e modes, but we shall not pursue this matter further here.

\acknowledgments
The authors are grateful to Professor Lorenzo Marrucci for useful correspondence. 
\bibliography{MyBib}
\end{document}